\documentclass[aps,pra,twocolumn]{revtex4-1}
\usepackage{amsmath,amssymb,graphicx}
\usepackage[T1]{fontenc}

\begin{document}

\newcommand{\bra}[1]    {\langle #1|}
\newcommand{\ket}[1]    {|#1 \rangle}
\newcommand{\braket}[2]    {\langle #1 | #2 \rangle}
\newcommand{\ketbra}[2]{|#1\rangle\!\langle#2|}
\newcommand{\tr}[1]    {{\rm Tr}\left[ #1 \right]}
\newcommand{\av}[1]    {\left\langle #1 \right\rangle}
\newcommand{\modsq}[1]{\left|#1\right|^2}
\newcommand{\bflambda}{\boldsymbol{\lambda}}
\newcommand{\en}{\mathcal E_N}
\newcommand{\enn}[1]{\mathcal E_{#1}}

\title{Characterizing quantum correlations in spin chains}
\author{Artur Niezgoda, Mi{\l}osz Panfil and Jan Chwede\'nczuk}
\affiliation{Faculty of Physics, University of Warsaw, ul. Pasteura 5, PL--02--093 Warszawa, Poland}
\begin{abstract}
  The growth in the demand for precisely crafted many-body systems of spin-$1/2$ particles/qubits is due to their top-notch versatility in application-oriented quantum-enhanced protocols and
  the fundamental tests of quantum theory. Here we address the question: how quantum is a chain of spins? We demonstrate that a single element of the density matrix carries the answer. Properly
  analyzed it brings information about the extent of the many-body entanglement and the non-locality. 
  This method can be used to tailor and 
  witness highly non-classical effects in many-body systems with possible applications to quantum computing, ultra-precise metrology or large-scale tests of quantum mechanics. 
  As a proof of principle, we investigate the extend of non-locality and entanglement in ground states and thermal states of experimentally accessible spin chains.
\end{abstract}
\maketitle

\section{Introduction}

An ensemble of spin-$1/2$ particles is a paradigm of a complex many-body quantum system---an ideal probe of various aspects of the theory, ranging from quantum phase transitions~\cite{SachdevBOOK} to many-body entanglement
~\cite{giovannetti2004quantum} or the non-locality~\cite{Tura1256}.
Correlated states of many qubits are at the core of quantum-enhanced metrology~\cite{giovannetti2004quantum,pezze2018quantum}, 
quantum-information processing~\cite{monroe2002quantum} and tests of foundations of quantum mechanics. The quantum information aspects play also an increasingly important role in the condensed matter physics~\cite{QI_CondMatBOOK}. 
The experimental advances in the field of quantum simulators made it possible to prepare, control and measure 
with a great precision quantum many-body states~\cite{2014RvMP...86..153G}. The expanding toolbox includes
ultra-cold atoms~\cite{esteve2008squeezing,riedel2010atom,gross2010nonlinear,leroux2010orientation,smerzi_ob,Bloch:2012aa,Hofstetter_2018,Browaeys:2020aa}, 
trapped ions~\cite{ion1,ion2,haffner2005scalable,blatt2008entangled,Korenblit_2012,Blatt:2012aa,Bohnet1297} and super-conducting qubits~\cite{Houck:2012aa,super1,super2} among others. 

Given the growing interest  it is relevant to adequately characterize the quantum features of a multi-qubit state. What can we learn about the
system given a value of some correlation function? Is it really quantum or can it be reproduced with some rather classical ensemble? Here we address these questions in a systematic way relevant for experimentally realisable systems of dozens of spins.
Many powerful measures of entanglement, like the entanglement entropy~\cite{QI_CondMatBOOK,Calabrese_2004,Hastings_2007,Franchini_2007,PhysRevB.81.060411,Alba_2012,LAFLORENCIE20161,Wald_2020} or negativity~\cite{PhysRevA.58.883,Calabrese_2013,Ruggiero_2016}, require detailed knowledge of the density matrix which makes 
their experimental measurements challenging~\cite{Islam:2015aa,Hauke:2016aa}. 
In this work we propose an alternative approach. 
We show that a single element of the density matrix---related to the formation probability~\cite{KorepinBOOK,Shiroishi_2001,Razumov_2001,2003PhLA..316..342A}---
carries precise information about the many-body entanglement~\cite{ent_rmp} and the ultra-quantum Bell non-locality~\cite{epr,bell,bell_local}.
The experimental implementation of the devised protocol requires single-atom resolved spectroscopy which is within the range of experimental techniques~\cite{sherson2010single} and should allow for testing entanglement and non-locality in correlated systems. 

In this work we focus on the
paradigmatic and experimentally accessible quantum Ising model~\cite{Kim_2011,Simon:2011aa,Bernien_2017,de_L_s_leuc_2018}, the XXZ spin chain~\cite{PhysRevLett.115.215301,Gras:2014aa}
and the Majumdar-Ghosh model~\cite{Majumdar_Ghosh,Majumdar_1970}. Our aim is in understanding the extend of entanglement and non-locality in the groundstates of short chains. For example, to distinguish the situation in which the $6$-spin ground state of the Ising chain is actually a state where the entanglement/non-locality extends over 
all spins ($6$-partite entanglement~\cite{hyllus2012fisher,toth2012multipartite}), from a state 
in which the mutual quantum correlation extends only over $4$ spins. 
Besides enriching our understanding of quantum correlated states of matter, such information should be useful in optimisation 
of numerical approaches like Density Matrix Renormalisation Groups~\cite{PhysRevLett.69.2863,2005_Schollwock_RMP_77} and related methods~\cite{Schollw_ck_2011,Or_s_2019}.

The quantum information backbone of our work relies on a
class of Bell inequalities, which make no assumption on the number of 
particles/subsystems or on how the local outcomes are bounded, introduced by Cavalcanti and collaborators~\cite{cavalcanti2007bell} (see also~\cite{he2011entanglement,cavalcanti2011unified}).
We briefly review the derivation of this inequality in Section~\ref{sec.der}, which can be used to detect both the entanglement and non-locality. 
We also show an important property of this correlator---it extracts the
entanglement and the non-locality from a single element of the density matrix.
Next, we show that the value of this element allows to track how the entanglement/non-locality
spread over the many-body system.
Though experimental measurements of high-order
correlation functions are difficult, recent years fruited in a number of detection schemes with an efficiency at the level of a 
single atom/ion~\cite{bucker2009single,sherson2010single,dall2013ideal,shin2019bell,lopes2015atomic}. Therefore, the hierarchical method for testing
entanglement/non-locality in systems of growing complexity could be gradually tested in coming experiments. 
In Section~\ref{sec.exa} we apply this tool to analyze the build-up of many-body entanglement/non-locality in a spin chain described by various physical Hamiltonians, both in the ground and
thermal state. Finally, we conclude in Section~\ref{sec.conc}. Some details of the analytical calculations are presented in the Appendices.

\section{The Bell inequality for $N$ qubits}\label{sec.der}

We consider a system composed of $N$ parts. Measurements of each yield two binary outcomes $\sigma^{(k)}_x=\pm1$ and $\sigma^{(k)}_y=\pm1$ (with $k=1\ldots N$). 
We introduce a correlator
\begin{align}\label{eq.e}
  \mathcal C_N=\av{\sigma^{(1)}\cdot\ldots \sigma^{(N)}},
\end{align}
where $\sigma^{(k)}=\frac12(\sigma_x^{(k)}+i\sigma_y^{(k)})$. The ``$+$'' sign here can be changed to ``$-$'' independently for each party and the arguments that follow hold.
If the above mean can be reproduced with a probability distribution $p(\lambda)$ of a random (hidden) variable
$\lambda$, that correlates the outcomes in a classical way, i.e.,
\begin{align}
  \mathcal C_N=\int\!d\lambda\,p(\lambda)\sigma^{(1)}(\lambda)\cdot\ldots\cdot\sigma^{(N)}(\lambda),
\end{align}
then the correlator is consistent with a local hidden-variable theory (LHV).
Using a Cauchy-Schwarz inequality (CSI) we obtain the following  Bell inequality~\cite{cavalcanti2007bell,he2010bell} for $\en=\modsq{\mathcal C_N}$:
\begin{align}\label{eq.bell.ineq}
  \en\leqslant\int\!d\lambda\,p(\lambda)|\sigma^{(1)}(\lambda)|^2\cdot\ldots\cdot|\sigma^{(N)}(\lambda)|^2=2^{-N},
\end{align}
where the last step is a consequence of $\modsq{\sigma^{(k)}(\lambda)}=\frac12$.
If we consider quantum-mechanical systems, then $\sigma^{(k)}(\lambda)$'s are replaced by the Pauli rising operators for each qubit/spin-$1/2$ particle. 
The Bell inequality reads
\begin{align}\label{eq.b.op}
  \en=\modsq{\av{\bigotimes_{k=1}^N\hat\sigma_+^{(k)}}}\leqslant2^{-N}.
\end{align}
The breaking of this inequality proofs that spins form a non-local state.
If we take a separable state of $N$ qubits
\begin{align}\label{eq.sep}
  \hat\varrho_N=\int\! d\lambda\,p(\lambda)\,\bigotimes_{k=1}^N\hat\varrho^{(k)}(\lambda),
\end{align}
then using $\tr{\hat\varrho^{(k)}(\lambda)\hat\sigma_+^{(k)}}=\modsq{\av{\hat\sigma^{(k)}_+}}_\lambda\leqslant\frac14$ we obtain another, less restrictive bound
\begin{align}\label{eq.ineq.ent}
  \en\leqslant\int\! d\lambda\,p(\lambda)\prod_{k=1}^N\modsq{\av{\hat\sigma^{(k)}_+}}_\lambda\leqslant2^{-2N},
\end{align}
breaking of which signals that qubits form a non-separable state~\cite{cavalcanti2011unified}.

Note that, according to~\eqref{eq.b.op}, $\en=|\varrho_{a,b}|^2$, where  $\varrho_{a,b}$ is a component of the density matrix that couples
$\ket{\psi_a}\equiv\ket{\uparrow_1,\ldots,\uparrow_N}$ with
$\ket{\psi_b}\equiv\ket{\downarrow_1,\ldots,\downarrow_N}$, while $\hat\sigma_+^{(k)}\ket{\downarrow_k}=\ket{\uparrow_k}$.
This is an important observation---the $\en$ extracts entanglement and non-locality from a single element of the density matrix. Moreover, this leads to a size-independent upper bound
\begin{align}\label{eq.b.dens}
  \en=|\varrho_{a,b}|^2\leqslant\varrho_{a,a}\varrho_{b,b}\leqslant\frac14,
\end{align} 
which
implies that the inequality~\eqref{eq.ineq.ent} can be violated starting already from $N=2$, while the inequality~\eqref{eq.b.op} from $N=3$. Both are saturated by the maximally entangled GHZ state
\begin{align}\label{eq.st}
  \ket\psi=\frac1{\sqrt2}\left(\ket\uparrow^{\otimes N}+\ket\downarrow^{\otimes N}\right).
\end{align}
We now argue that the value of $\en$ carries detailed information on the multiparticle entanglement and non-locality. We illustrate this with three densities matrices of different character.

\subsection{Many-body entanglement and non-locality}\label{sec.hierarchy}

In the first example we consider a system, where out of $N$ spins, two form an entangled state, and the other $N-2$ are separable, i.e., 
\begin{align}\label{eq.ent.2}
  \hat\varrho_N=\int\! d\lambda\,p(\lambda)\,\left(\bigotimes_{k=1}^{N-2}\hat\varrho^{(k)}(\lambda)\right)\otimes\hat\varrho_2(\lambda).
\end{align}
(The lower index of the density matrix is the number of spins it describes, while the upper index $(k)$ labels a single $k$-th spin.)
For this density matrix, the correlator $\en$ can be bounded using the CSI as follows
\begin{align}\label{eq.ent.2.ineq}
  \en&\leqslant\int\! d\lambda\,p(\lambda)\,\mathcal E_{N-2}(\lambda)\int\! d\lambda\,p(\lambda)\,\mathcal E_2(\lambda)\nonumber\\
  &\leqslant4^{-(N-1)},
\end{align}
where $\mathcal E_{N-2}(\lambda)$ is calculated with the product state of $N-2$ spins,  $\bigotimes_{k=1}^{N-2}\hat\varrho^{(k)}(\lambda)$, and
$\mathcal E_2(\lambda)$ is the two-spin correlator calculated with the density matrix $\hat\varrho_2(\lambda)$.
In the last step we used the upper bound from Eq.~\eqref{eq.b.dens}. Inequality~\eqref{eq.ent.2.ineq} is saturated by a product
of a two-spin GHZ state~\eqref{eq.st} and $N-2$ states
\begin{align}
  \ket{\psi_k}=\frac1{\sqrt2}\left(\ket{\!\uparrow_k}+e^{i\varphi_k}\ket{\!\downarrow_k}\right),
\end{align}
where $\varphi_k$ is an arbitrary phase. The violation of the bound~\eqref{eq.ent.2.ineq}, more stringent than~\eqref{eq.ineq.ent}, signals that
the entanglement extends either over more pairs than just one or over more than a pair (three-spin entanglement or more). 

As a second example, we analyse the state with pair-wise entangled spins, which reads
\begin{align}
  \hat\varrho_N=\int\! d\lambda\,p(\lambda)\,\bigotimes_{k=1}^{\frac N2}\hat\varrho_2^{(k)}(\lambda),
\end{align}
where $\hat\varrho_2^{(k)}(\lambda)$ is a density operator of the $k$-th pair (we took $N$ even for simplicity). In this case
$\en\leqslant4^{-\frac N2}$ and if violated, the state is at least three-spin entangled. 

Finally, if all-but one spin form a $N-1$ entangled state, separable with the $N$-th spin,
\begin{align}
  \hat\varrho_N=\int\! d\lambda\,p(\lambda)\,\hat\varrho_{N-1}(\lambda)\otimes\hat\varrho^{(N)}(\lambda),
\end{align}
then $\en<\frac{1}{16}$. Values of the correlator from the range $]\frac1{16},\frac14]$ would imply then the $N$-particle entanglement.

Similarly, the value of $\en$ brings information about the extend of non-local correlations in the spin system. 
For instance, when the correlation among no more than
three out of $N$ spins cannot be explained with a LHV theory, then, in analogy to Eq.~\eqref{eq.ent.2.ineq}
\begin{align}\label{eq.loc.2}
  \en&\leqslant\int\! d\lambda\,p(\lambda)\,\mathcal E_{N-3}(\lambda)\int\! d\lambda\,p(\lambda)\,\mathcal E_3(\lambda)\leqslant2^{-(N-1)}.
\end{align}
Here we used the fact that for the locally correlated $N-3$ spins, $\mathcal E_{N-3}(\lambda)\leqslant 2^{-(N-3)}$, while $\mathcal E_3(\lambda)\leqslant2^{-2}$.
When the non-local correlation extends over $N-1$ spins but not over the $N$-th, then $\en\leqslant\frac18$. Similarly to the entanglement witness, values
$\en\in]\frac18,\frac14]$ are accessible only to systems where the non-locality encompasses all the spins.

\section{Application to many-body physics} \label{sec.exa}

We now illustrate these general considerations with some prominent examples of spin-chain systems. In this context, the correlators of type $\enn{m}$ are known as formation probabilities (FP)
\begin{align}\label{eq.anti}
  \enn{m}=\modsq{\av{\hat\sigma_{\pm}^{(1)}\otimes\hat\sigma_{\pm}^{(2)}\otimes\hat\sigma_{\pm}^{(3)}\ldots \otimes\hat\sigma_{\pm}^{(m)}}},
\end{align} 
where the expectation value is computed in a system consisting of $N$ spins, and the average is taken, for example, in a ground state of some Hamiltonian. Formation probabilities in spin-chains have been studied for some time~\cite{KorepinBOOK,Shiroishi_2001,Razumov_2001,2003PhLA..316..342A, 10.1007/1-4020-4531-X_5,2005cond.mat..4307A,2009arXiv0904.3519S, Rajabpour_2015,10.1088/1751-8121/ab8507,Najafi_2020} with the focus on behaviour of $\enn{m}$ in the thermodynamically large system. Here, instead we focus on finite systems and, even more importantly, we build upon the hierarchy introduced in the previous section to develop detailed ''tomography'' of the entanglement and non-locality in the ground states of experimentally relevant Hamiltonians.

The choice of the signs in~\eqref{eq.anti} should maximize the correlator and can be motivated by the expected structure of correlations in the considered state. In practice, 
the adequate correlator can be chosen by looking at the largest off-diagonal value of the density matrix in the local spin basis, with {\rm e.g.} exact diagonalization.
\begin{figure}[t!]
  \includegraphics[width=\columnwidth]{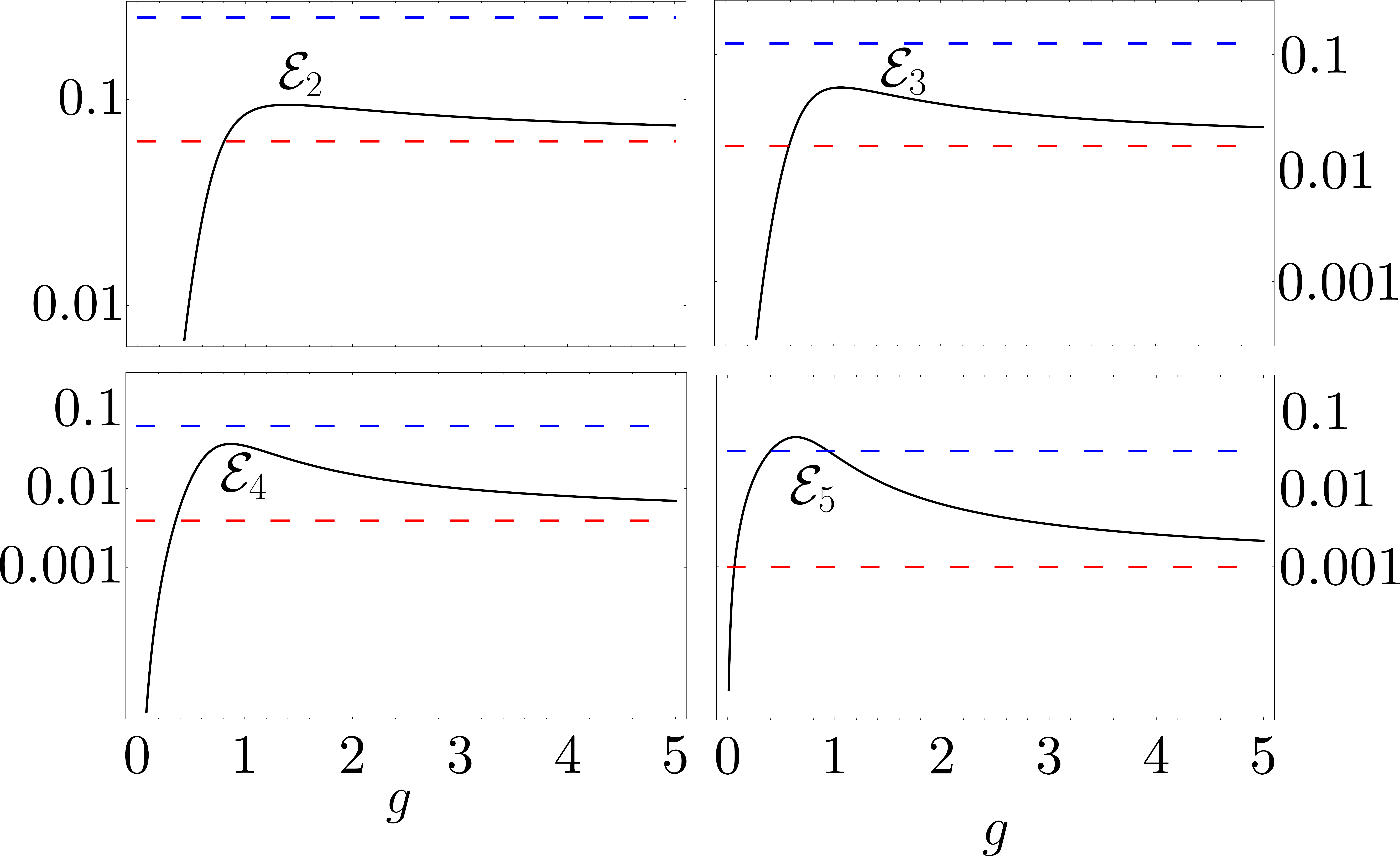}
  \caption
      {Correlators $\enn m$ for $m=2\ldots5$ as a function of $g$ calculated with the ground state of the Ising Hamiltonian~\eqref{eq.ham.ising} for $N=6$. 
        The horizontal dashed lines denote the entanglement bound $4^{-m}$ (red) and the non-locality bound $2^{-m}$ (blue).}
      \label{fig.corr}
\end{figure}

We start with the Ising Hamiltonian (with open boundary conditions) in the antiferromagnetic phase
\begin{align}\label{eq.ham.ising}
  \hat H=\sum_{j=1}^{N-1}\hat\sigma_z^{(j)}\hat\sigma_z^{(j+1)}+g\sum_{j=1}^N\hat\sigma_x^{(j)}.
\end{align}
Here,  the control parameter $g$ is the magnitude of the external magnetic field in the $x$-direction. We solve the model numerically by doing the exact diagonalization of the Hamiltonian for $N$=6.
We adjust the form of the correlator from Eq.~\eqref{eq.b.op} to detect the quantum properties in this aniferromagnetic phase. This is done
by taking the rising/lowering operators alternating from site to site, 
\begin{align}\label{eq.anti}
  \enn{m}=\modsq{\av{\hat\sigma_+^{(1)}\otimes\hat\sigma_-^{(2)}\otimes\hat\sigma_+^{(3)}\ldots \otimes\hat\sigma_{\pm}^{(m)}}}.
\end{align}
with five possibilities: $m\in[2,6]$. Fig.~\ref{fig.corr} shows the first four correlators as a function of $g\in[0,5]$ with entanglement ($4^{-m}$) and non-locality ($2^{-m}$) bounds marked. 
While the appearance of entanglement is witnessed by $\enn m$ starting from the lowest order $m=2$, the Bell correlations are detected only at $m=5$. All the correlators drop to zero as $g\rightarrow 0$. This
is because for the vanishing magnetic field, the ground state is a superposition of two anti-ferromagentic states
\begin{align}\label{eq.g0}
  \ket\psi=\frac1{\sqrt2}\left(\ket{\!\uparrow\downarrow,\ldots}+\ket{\!\downarrow\uparrow,\ldots}\right).
\end{align}
Tracing out a single spin from such a maximally entangled state gives a classical mixture,  with no quantum features.
When $g\gg1$, the ground state is that of the non-interacting
spins, which explains why all the correlators drop as $g$ grows and ultimately tend to $4^{-m}$. 
We also observe that the Bell correlations are persistent around the critical point $g=1$ of the quantum phase transition in quantum Ising chain. In the region of a phase transition the correlation 
length is large (approaching infinity in the thermodynamically large system). In such situation tracing out part of the system does not destroy the correlations within the subsystem what allows 
for $\enn{m}$ to stay large. This analysis suggests the hierarchy of $\enn{m}$'s is an appropriate tool for exploring entanglement and non-locality 
around the quantum phase transitions. 

\begin{figure}[t!]
  \includegraphics[width=\columnwidth]{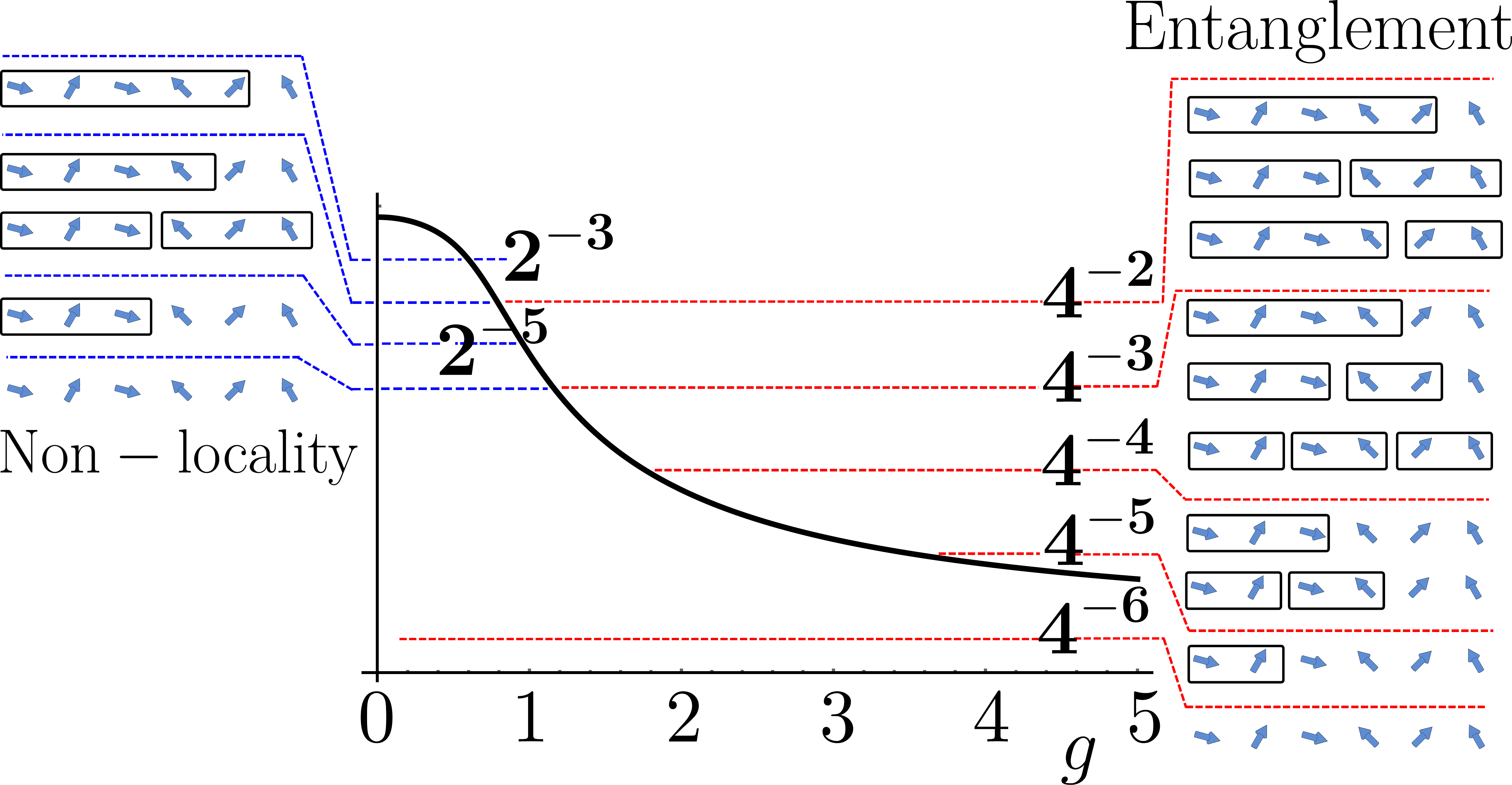}
  \caption
      {The full six-spin correlator $\enn6$ (solid black line) calculated with the groundstate of the Hamiltonian~\eqref{eq.ham.ising} as a function of $g$. The horizontal red 
        (entanglement) and blue (non-locality)  dashed lines separate regions, where $\enn6$ can be reproduced with a spin system with a specific multiparticle correlation, 
        see the main text for an explanation. }
      \label{fig.e6}
\end{figure}

The $m=6$ case is shown separately in Fig.~\ref{fig.e6}. The $\enn6$ reaches its maximal value for $g=0$ as it detects the quantum properties of the fully entangled state~\eqref{eq.g0}.
We mark not only the entanglement- and the non-locality-lower bound ($4^{-6}$ and $2^{-6}$) but also all the other limits derived from
the considerations introduced in Eqs~\eqref{eq.ent.2}-\eqref{eq.loc.2}. This figure should be read as follows. When $\enn6<4^{-6}$, the correlation can be reproduced with a separable state of 6 spins
(all six arrows on the RHS of the plot unboxed). When $\enn6\in]4^{-6},4^{-5}[$, the correlation can be reproduced with a setup, where two spins are entangled and other four form a 
separable state (2x1x1x1x1: two spins in a box, other unboxed). This is the most classical two-spin entangled state. Higher two-spin entangled states can be used to explain the correlation strength in 
the range $\enn6\in]4^{-5},4^{-4}[$ (2x2x1x1) or $\enn6\in]4^{-4},4^{-3}[$ (2x2x2). All other cases are visually shown on the RHS of Fig.~\ref{fig.e6}. Similarly, when $\enn6<2^{-6}$, the correlator
can be modelled with a full LHV theory. When $\enn6\in]2^{-6},2^{-5}[$, at least two spins are non-locally correlated, and so forth.

\subsubsection{The Ising model with long range interactions}
\begin{figure}[t!]
  \includegraphics[width=\columnwidth]{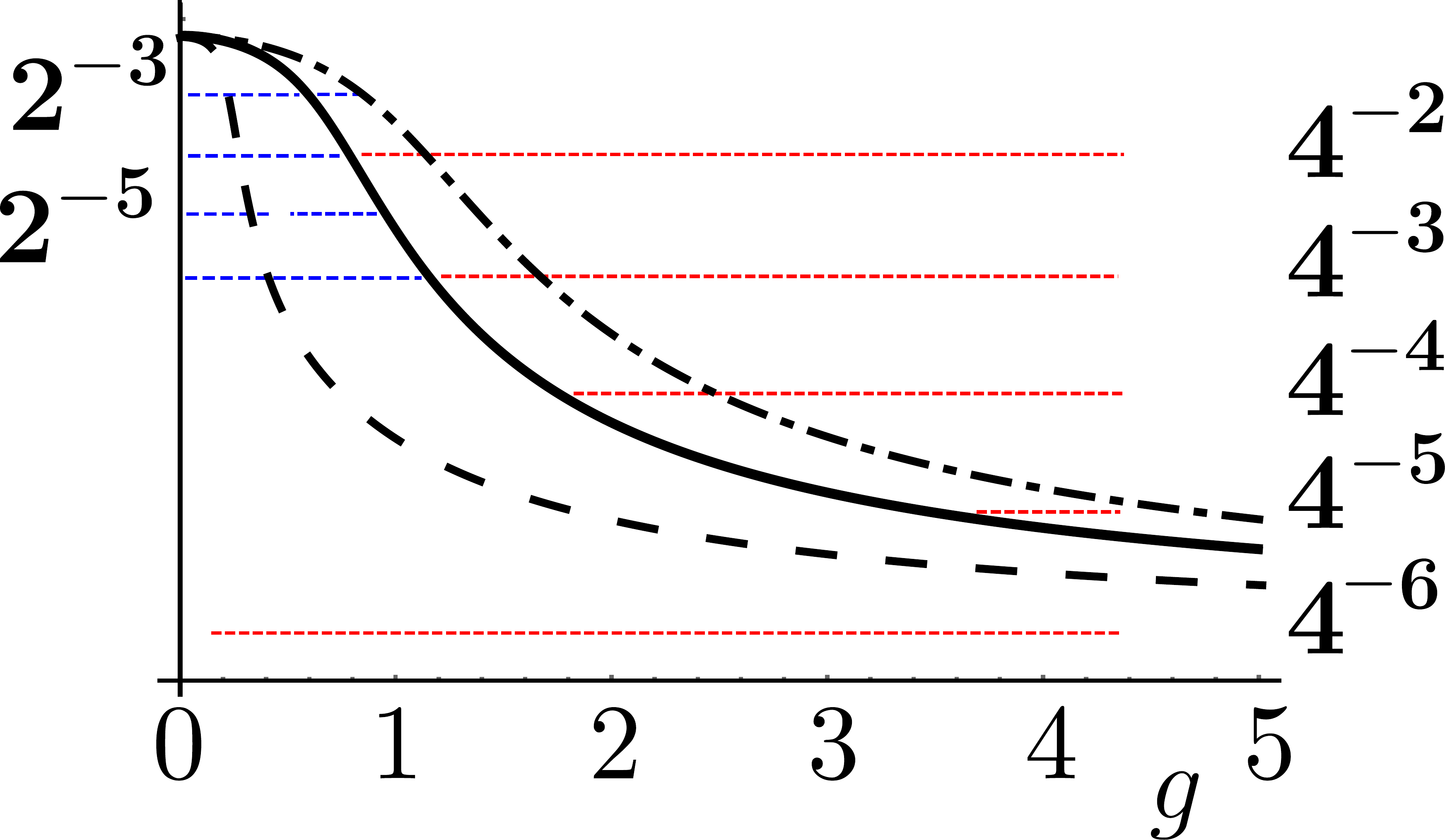}
  \caption{The correlator $\enn6$ calculated with the Hamiltonian~\eqref{eq.ising.long} for $K=0$ (solid black line), $K=0.4$ (dashed) and $K=-0.4$ (dot-dashed).}
  \label{fig.ising.long}
\end{figure}

\begin{figure}[t!]
	\includegraphics[width=\columnwidth]{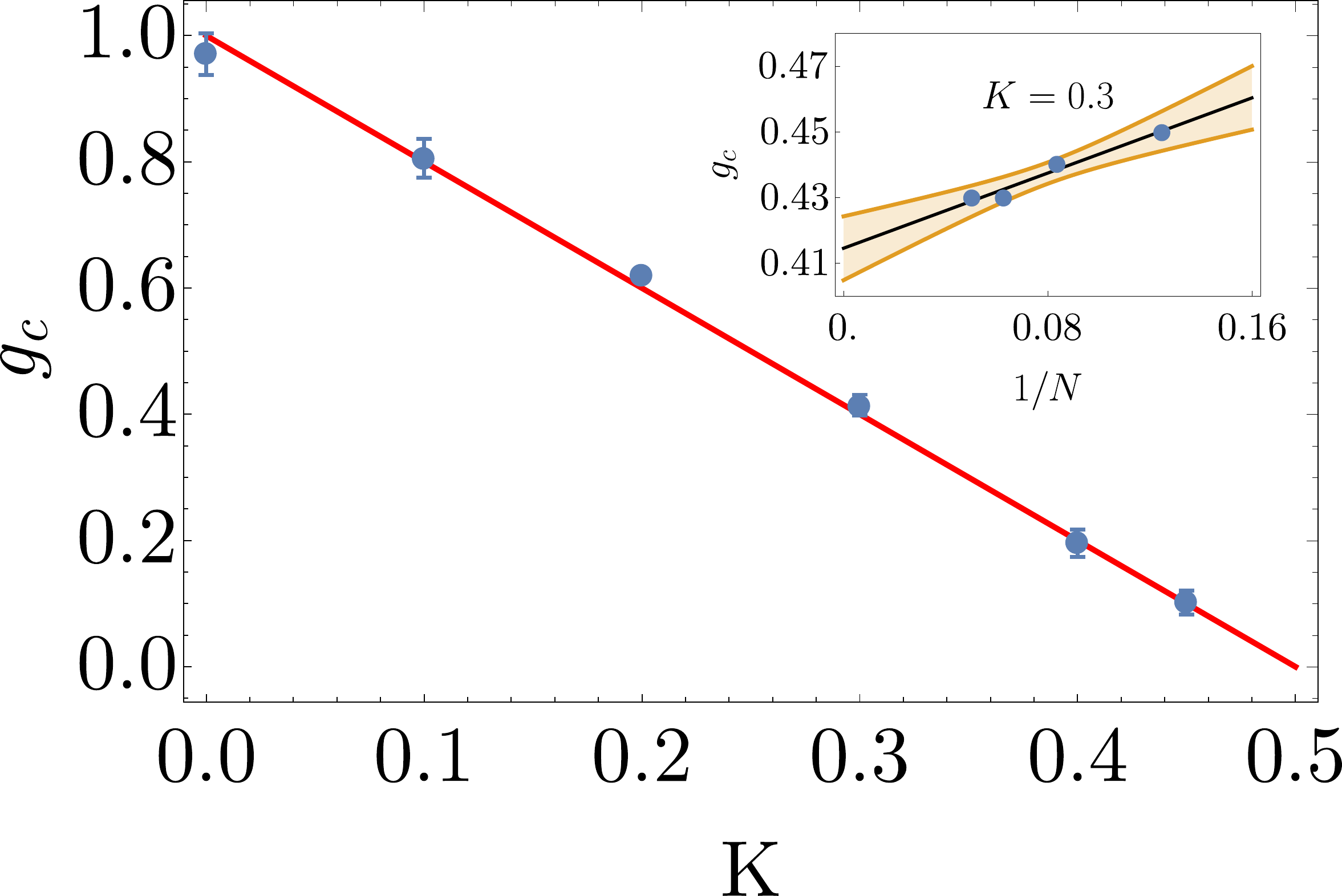}
	\caption{Blue dots are results of finite size scaling of the position of the maximum of $\enn{N/2}$
          using $N = 8, 12, 16, 20$
          (an inset shows an example of scaling for $K=0.3$). The red line denotes the quantum phase transition separating the antiferromagnetic phase from paramagnetic~\cite{PhysRevB.76.094410,PhysRevE.75.021105,Nagy_2011}. The error bars comes from the linear fit estimation of the finite size scaling. The shaded region in the inset represents the $0.9$ confidence interval.}
	\label{fig:scaling}
\end{figure}

We now test the impact of longer-range interactions on the quantum correlations. To this end, we include a term which connects every spin with the next-to-adjacent one,
\begin{align}\label{eq.ising.long}
  \hat H=\sum_{j=1}^{N-1}\hat\sigma_z^{(j)}\hat\sigma_z^{(j+1)}+g\sum_{j=1}^N\hat\sigma_x^{(j)}+K\sum_{j=1}^{N-2}\hat\sigma_z^{(j)}\hat\sigma_z^{(j+2)}.
\end{align}
We expect that turning on ferromagnetic ($K<0$) interaction should strengthen the entanglement in the ordered ($g<1$) phase and weaken it if the interaction is antiferromagnetic ($K>0)$ and competes with the nearest neighbour interaction. 
Fig.~\ref{fig.ising.long} compares the $\enn6$ for $K=0$ and $K=\pm0.4$ and confirms these expectations.
Turning on the $K$ parameter allows us also to study more deeply the relation between the hierarchy of the correlations and the quantum phase transition. The position of the critical point, for the competing interaction ($K >0$), was studied before~\cite{PhysRevB.76.094410,PhysRevE.75.021105,Nagy_2011} with the conclusion that for intermediate values of $K < 0.5$ the critical point is given by $g_c = 1 - 2 K$. From Fig.~\ref{fig.corr} we observe that for the system size $N=6$, the correlator $\enn3$ has a maximal value close to the critical point $g_c=1$ of a short range Ising model. Guided by this observation we perform the finite-size scaling of the position of the maximum of $\enn{N/2}$ for chains of length $N = 8, 12, 16, 20$ and for different values of $K$. The results are presented in Fig.~\ref{fig:scaling} and show that maximum of $\enn{N/2}$ coincides with the position of the phase transition. We note however, that $\enn{N/2}$ does not seem to exhibit a singularity at the critical point and therefore is not a standard order parameter~\cite{SachdevBOOK}.

The results for the Ising models show how the hierarchy of the correlation can be exploited to understand the quantum features of the many-body ground states extending from the detailed tomography of entanglement and non-locality 
to the detection of quantum phase transitions. In the remainder of the paper we will examine two more models, the XXZ spin chain and the Majumdar-Ghosh model both enjoying analytical computation of the correlations. In the former case we will explore the effect of the finite temperature on the hierarchy and show that the correlations are naturally expressible in the language of the Bethe Ansatz solution. In the latter, the valence bound structure of its ground state brings intuitive understanding when to expect strong correlations.

\subsection{The XXZ quantum spin chains}
As another emblematic spin-chain system we consider the XXZ model~\cite{KorepinBOOK} with the Hamiltonian
\begin{align}\label{XXZ_H}
  \hat H =\sum_{j=1}^N(\hat{\sigma}^{(j)}_x\hat{\sigma}^{(j+1)}_x + \hat{\sigma}^{(j)}_y\hat{\sigma}^{(j+1)}_y + \Delta\hat{\sigma}^{(j)}_z\hat{\sigma}^{(j+1)}_z),
\end{align}
where $\Delta$ is the anisotropy parameter and we assume the periodic boundary conditions. $\Delta =1$ corresponds to the $SU(2)$ symmetric XXX model while for $\Delta = 0$ it simplifies to the 
XX Hamiltonian, equivalent to free fermions. Finally,  for $\Delta \to \pm\infty$ one obtains the Ising model at zero magnetic field. The XXZ Hamiltonian exhibits quantum phase transitions at $\Delta = \pm 1$ separating a ferromagnetic phase ($\Delta < -1$) from paramagnetic ($|\Delta| < 1$) and antiferromagnetic ($\Delta > 1)$. We focus on the regime with $\Delta > -1$ and start with analysing an exact solution with $N=4$.

\subsubsection{4-spin chain}
\begin{figure}[t!]
  \center
  \includegraphics[width=\columnwidth]{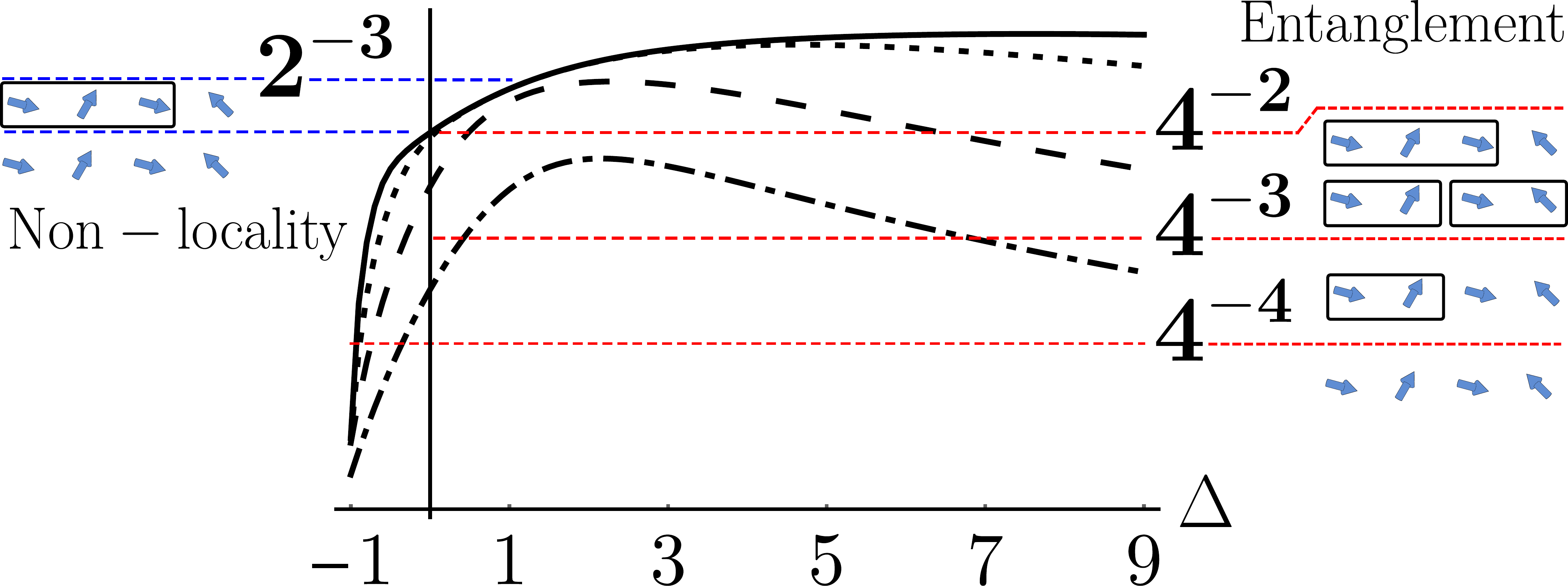} 
  \caption{$\enn4$ from Eq.~\eqref{eq.b4} as a function of $\Delta$ for $\beta=10$ (solid), $\beta=5$ (dotted), $\beta=2$ (dashed) and $\beta=1$ (dot-dashed). }
  \label{fig.beta}
\end{figure}

The XXZ Hamiltonian is exactly solvable by the Bethe Ansatz methods for any $N$. Particularly simple solution, not relying on the Bethe Ansatz technique, exists for $N=4$. We will 
analyze this special case as it allows to easily include thermal effects.
Using the thermal density matrix
\begin{align}
  \hat{\varrho}_T=\frac1{\mathcal{Z}}\sum_ne^{-\beta E_n}\ketbra{\psi^{(n)}}{\psi^{(n)}},
\end{align}
where the summation runs through all the eigen-levels ($\hat H\ket{\psi^{(n)}}=E_n\ket{\psi^{(n)}}$) of the $4$-spin Hamiltonian from Eq.~\eqref{XXZ_H} and $\beta=(k_BT)^{-1}$, [$T$ is the
temperature, $k_B$ the Boltzmann constant and $\mathcal{Z}$ is the statistical sum]. The spectrum consists of $16$ states out which $3$ have non-zero expectation values
\begin{subequations}
  \begin{align}
    \left|\langle E_{\pm} | \hat\sigma^{(1)}_+\hat\sigma^{(2)}_-\hat\sigma^{(3)}_+\hat\sigma^{(4)}_- |E_{\pm} \rangle \right|&=\frac14  \left| 1 \mp \frac{\Delta}{\sqrt{8 + \Delta^2}} \right|,\\
    \left|\langle E_{\Delta}| \hat\sigma^{(1)}_+ \hat\sigma^{(2)}_- \hat\sigma^{(3)}_+ \hat\sigma^{(4)}_- |E_{\Delta}\rangle\right|&=\frac12,
  \end{align}
\end{subequations}
where $\hat H |E_{\pm} \rangle = E_{\pm} |E_{\pm}\rangle$ and $\hat H |E_{\Delta}\rangle = -\Delta |E_{\Delta}\rangle$ with
\begin{equation}
  E_{\pm} = \frac{1}{2}\left(-\Delta \pm \sqrt{8 + \Delta^2} \right),
\end{equation} 
and $|E_-\rangle$ being the ground state for $\Delta > - 1$. In the local spin basis these states are 
\begin{align}
  |E_{\pm} \rangle &= \mathcal{N_{\pm}} \left(\frac{\Delta \pm \sqrt{8 + \Delta^2}}{2\sqrt{2}} |AF_2\rangle + |AF\rangle\right) , \\
  |E_{\Delta} \rangle &= \frac{1}{\sqrt{2}} \left( |\!\uparrow \downarrow \uparrow \downarrow \rangle - |\!\downarrow \uparrow \downarrow \uparrow\rangle\right),
\end{align}
with
\begin{align}
  |AF_2\rangle &= \frac{1}{2} \left( |\!\uparrow \uparrow \downarrow \downarrow\rangle + |\!\downarrow \uparrow \uparrow \downarrow\rangle + |\!\downarrow \downarrow \uparrow \uparrow\rangle + |\! \uparrow \downarrow \downarrow \uparrow \rangle \right), \\
  |AF\rangle &= \frac{1}{\sqrt{2}} \left( |\!\uparrow \downarrow \uparrow \downarrow \rangle + |\!\downarrow \uparrow \downarrow \uparrow\rangle\right),
\end{align}
and $\mathcal{N}_{\pm}$ appropriate normalization factors.
The partition function reads
\begin{subequations}\label{eq.b4}
  \begin{align}
    \enn4&=\frac{1}{\mathcal{Z}^2}\Bigg[-\frac{e^{-\beta(\Delta + \sqrt{8+\Delta^2})}}{2} +\frac{1}{4}\left(1 + \frac{\Delta}{\sqrt{8+\Delta^2}}\right)\nonumber\\
      &+\frac{1}{4}\left(1 - \frac{\Delta}{\sqrt{8+\Delta^2}}\right)e^{-2 \beta \sqrt{8+\Delta^2}} \Bigg]^2,\\
	\mathcal{Z}&= 1 + e^{-\beta(E_+ - E_-)} + e^{-\beta (-\Delta -E_-)} + 2 e^{-\beta (-1 - E_-)} \nonumber \\
	&+ 7 e^{\beta E_-} + 2 e^{-\beta(1 - E_-)} + 2e^{-\beta(\Delta - E_-)},
  \end{align}
\end{subequations}
for the anti-ferromagnetic product of four rising/lowering operators as in Eq.~\eqref{eq.anti}. This result 
is plotted in Fig.~\ref{fig.beta} for four values of $\beta$ as a function of $\Delta$. Since $\Delta$ sets the energy scale in this problem, we expect the Bell correlation to vanish for 
$\beta\lesssim\Delta$, as confirmed in the plot. Interestingly, even for quite large $\beta=2$, the correlation still cannot be modelled with a two-particle entangled state around $\Delta=1$.
In the limit of $T\rightarrow0$ we obtain
\begin{align}
  \enn4=\frac{1}{16}\left(1 + \frac{\Delta}{\sqrt{8+\Delta^2}}\right)^2.
\end{align}
This is bigger than the LHV limit $2^{-4}$ iff $\Delta>0$. It also crosses other entanglement/non-locality limits. For instance, when $\Delta>2\sqrt{\sqrt2-1}$, then $\enn4>2^{-3}$, which means
that the correlator can be reproduced only with four-spin entangled system where also all four spins are non-locally correlated.

These results shows that the hierarchy provides insights into a quantum-mechanical features of many-body systems also at finite temperatures.

\subsubsection{Bethe Ansatz solution}

\begin{figure}[t!]
	\center	
	\includegraphics[width=\columnwidth]{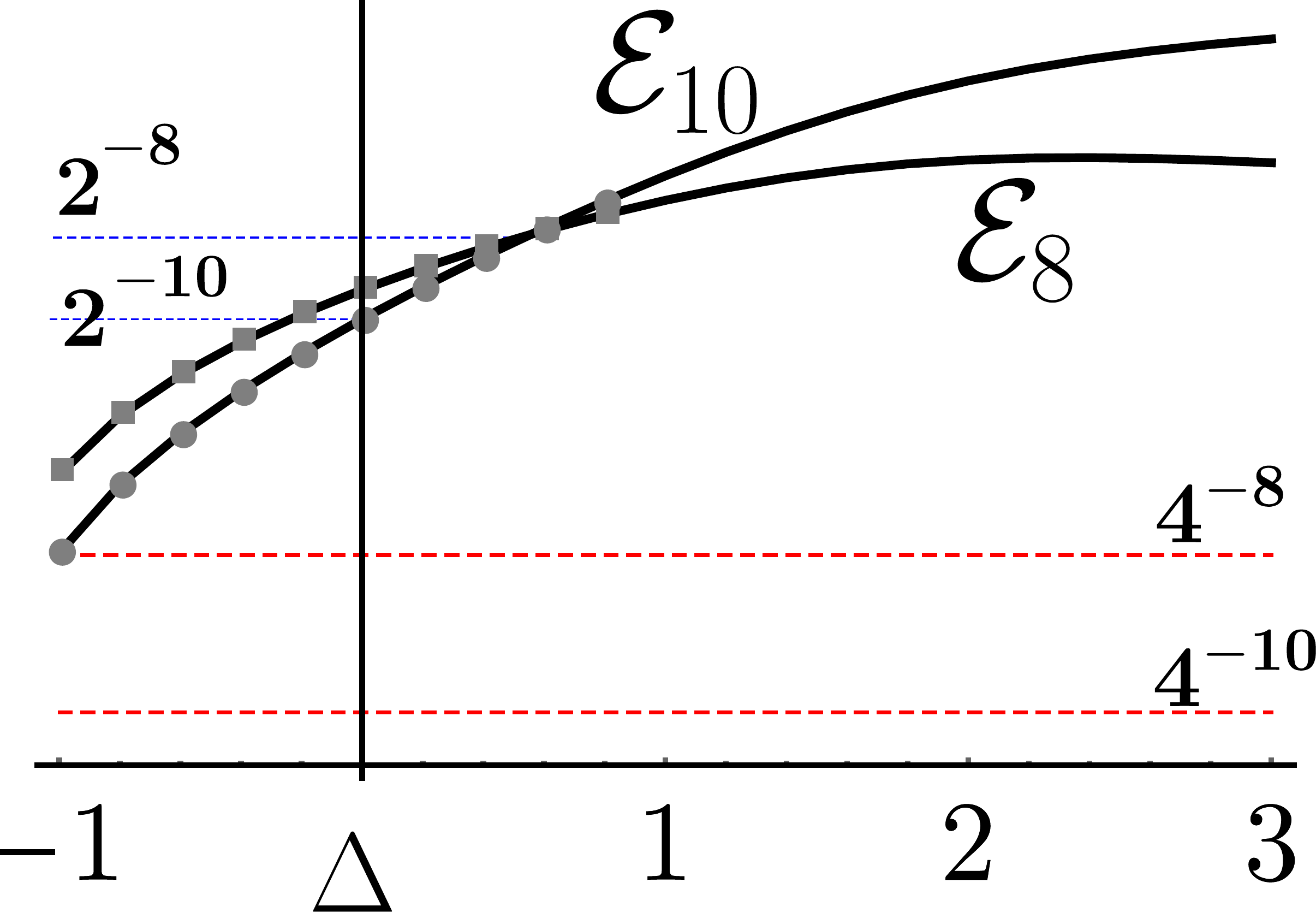}
	\caption{The ground state correlators $\enn{10}$ and $\enn{8}$ for the XXZ spin chain of length $N=10$ as a function of the anisotropy $\Delta$. 
          We show the analytic results (grey points) of Bethe Ansatz~\eqref{XXZ_E_N} and results of numerical diagonalization (black solid line) of the XXZ Hamiltonian~\eqref{XXZ_H}.}
	\label{fig:XXZ}
\end{figure}

The XXZ spin chain can be exactly solved with the Bethe Ansatz technique. In Appendix~\ref{app:XXZ} we recall the main ingredients of the solution that give us 
direct access to the correlator $\enn{N}$ in the ground state
\begin{equation}
	\enn{N} = \modsq{\chi({\rm O}_N|\bflambda_{N/2})}\cdot\modsq{\chi({\rm E}_N|\bflambda_{N/2})}, \label{XXZ_E_N}
\end{equation}
where $\chi({\rm E}_N|\bflambda_{N/2})$ and $\chi({\rm O}_N|\bflambda_{N/2})$ describe the antiferromagnetic components of the ground state wave-function. Lower order correlators are also accessible. The details of the notation are presented in Appendix~\ref{app:XXZ}. We also compute $\enn{m}$ with the help of exact diagonalization. Figure~\ref{fig:XXZ} shows the results for $N=10$, compared with the outcome of the numerical diagonalization of the Hamiltonian~\eqref{XXZ_H}.  

\subsection{Majumdar-Ghosh model}
Finally, we consider the Majumdar-Ghosh model~\cite{Majumdar_Ghosh,Majumdar_1970} with periodic boundary conditions, depicted by the Hamiltonian
\begin{align}
  \hat H=\sum_{\{j\}}\hat{\vec{\sigma}}^{(j)}\hat{\vec{\sigma}}^{(j+1)}+\frac12\sum_{\{j\}}\hat{\vec{\sigma}}^{(j)}\hat{\vec{\sigma}}^{(j+2)}.
\end{align}
Its (translationally invariant) ground state
\begin{align}
  \ket\psi=\mathcal N^{-1}\left(\ket{\psi_1} + \ket{\psi_2}\right)
\end{align}
is a superposition of two products of singlet states
\begin{subequations}
  \begin{align}
    &\ket{\psi_1}=\bigotimes_{j=0}^{\frac N2-1}\frac{\ket{\uparrow_{2j+1},\downarrow_{2j+2}}-\ket{\downarrow_{2j+1},\uparrow_{2j+2}}}{\sqrt2}\\
    &\ket{\psi_2}=\bigotimes_{j=0}^{\frac N2-1}\frac{\ket{\uparrow_{2j+2},\downarrow_{2j+3}}-\ket{\downarrow_{2j+2},\uparrow_{2j+3}}}{\sqrt2}.
  \end{align}
\end{subequations}
Here, $\mathcal N$ stands for the normalization. 
The explicit construction of the ground state for arbitrary $N$ allows for analytic computations of the antiferromagnetic correlator, as in Eq.~\eqref{eq.anti}. 
For $N/2$ even we obtain (for details, see Appendix~\ref{app:MG})
\begin{align}
  \en=\frac{1}{(1+2^{N/2-1})^2 },
\end{align}
which brakes the Bell limit $2^{-N}$ (and therefore also the entanglement limit $2^{-2N}$). The lower order correlations are also accessible. For $\enn{N-2}$ we find
\begin{equation}
	\enn{N-2} = \frac{1}{4} \enn{N}.
\end{equation}
We observe that breaking the Bell limit is solely because the ground state is the superposition of $|\psi_1\rangle$ and $|\psi_2\rangle$ states. Indeed, the $\enn{m}$ correlator on $|\psi_1\rangle$ is equal to $2^{-m}$ and it is the add-mixture of $|\psi_2\rangle$ that lifts it above the threshold. Thus the ground state of the Majumdar-Ghosh model is a simple example of a state for which the hierarchy of the correlators breaks the Bell limit.

\section{Conclusions and outlook}\label{sec.conc}
We have demonstrated that the multi-particle entanglement and the Bell non-locality in many-body systems of spin-$1/2$ particles can be traced-back to a single element of the density matrix.
We have shown that the value of this element contains information about the number of entangled/Bell-correlated spins. This allows to track how the quantum features
expand over a large distance and number of particles, for instance in spin-chains. This method could be used to tailor and 
witness highly non-classical effects in many-body systems with possible applications to quantum computing, ultra-precise metrology or large-scale tests of quantum mechanics. Furthermore, the observable, formation probability, is accessible experimentally with the current state-of-art in the field of quantum simulators. 

We have also shown that the lower order correlators detect a quantum phase transition: the critical value being correlated with the maxima of the correlators. Optimally, we would like to have a quantity that exhibits a singular behaviour at the phase transition. Whether such quantity can be constructed from $\enn{m}$ is an interesting open problem.

The results of our work create a new incentive to study formation probabilities and further extend existing techniques of their computations in the Bethe Ansatz models. Another 
interesting problem would the be computation of the thermodynamic limit of $\enn {N/2}$, given its relation to the quantum phase transition, with the techniques of the asymptotic expansion~\cite{2003PhLA..316..342A,2005cond.mat..4307A}.

\section{Acknowledgements} We are grateful to Vincenzo Alba and Fabio Franchini for careful reading of the manuscript and insightful comments. 
We would like to thank Pawe{\l} Jakubczyk and Krzysztof Wohlfeld for useful discussions.
AN and JC are supported by Project no. 2017/25/Z/ST2/03039, funded by the National Science Centre, Poland, under the QuantERA programme. MP acknowledges the support from the National Science Centre, Poland, under the SONATA grant~2018/31/D/ST3/03588.

\appendix

\section{Correlations in the XXZ spin chain} \label{app:XXZ}

Bethe Ansatz provides us with exact eigenstates of the system expressed as a superposition of states in the local spin basis~\cite{KorepinBOOK,Franchini_2017}
\begin{equation}
	|\Psi_M(\bflambda_M)\rangle =  \frac{1}{M!} \sum_{m_1, \dots, m_M=1}^N \chi(\mathbf{m}_M| \bflambda_M) | \mathbf{m}_M\rangle, \label{Bethe_state}
\end{equation}
where
\begin{equation}
	 |\mathbf{m}_M\rangle = \hat{\sigma}_-^{(m_1)} \cdots \hat{\sigma}_-^{(m_M)} |0\rangle_+,
\end{equation}
and $|0\rangle_+$ is the fully polarized state with all the spins up. Here bold symbol denotes set, $\bflambda_M = \{\lambda_j\}_{j=1}^M$. The amplitude $\chi(\mathbf{m}_M| \bflambda_M)$ is determined by the Bethe Ansatz methods and is parametrized by the rapidities $\bflambda_M$ solving the Bethe equations
\begin{equation}
	\theta_1(\lambda_j) = \frac{2\pi I_j}{N} - \frac{1}{N} \sum_{k=1}^M \theta_2(\lambda_j - \lambda_k), \quad j=1, \dots , M, \label{Bethe}
\end{equation}
The momentum and two-body scattering phase shift in the gappless phase ($|\Delta < 1|$) are
\begin{align}
	p(\lambda) &=  i \log \frac{\cosh(\lambda - i \eta)}{\cosh(\lambda + i \eta)}, \\
	\theta(\lambda) &= i \log \frac{\sinh(2 i \eta + \lambda)}{\sinh(2 i \eta - \lambda)}.
\end{align}
The quantum numbers $I_j$ in~\eqref{Bethe} for a ground state (in the zero external magnetic field) are $I_j^{\rm GS} = - \frac{M+1}{2} + j$ for $j=1, \dots, M$ with $M = N/2$. The amplitudes are
\begin{align} \label{Bethe_amplitude}
	\chi(\mathbf{m}_M|\bflambda_M) = \frac{1}{|\mathcal{N}_M|}&\sum_{\sigma \in \mathcal{P}_M} (-1)^{|\sigma|}  \exp\left( - i \sum_{j=1}^M m_j p(\lambda_{\sigma_j})\right) \nonumber \\
	&\times \exp\left(  - \frac{i}{2} \sum_{k>j} \theta(\lambda_{\sigma_k} - \lambda_{\sigma_j})\right).
\end{align}
where the normalization $\mathcal{N}_M$~is
\begin{equation}
	|\mathcal{N}_M|^2 = \frac{\det G_M}{\prod_{j=1}^M K_1(\lambda_j)},
\end{equation}
and guarantees that $\braket{\Psi_M}{\Psi_M} = 1$.  The factors appearing in the normalization are the Gaudin matrix 
\begin{equation}
	G_{jk} = \delta_{jk}\!\left(\! N K_{1}(\theta_j) - \sum_{m=1}^M K_2(\lambda_j - \lambda_m)\! \right) + K_2(\lambda_j - \lambda_k).
\end{equation}
and functions
\begin{align}
	K_1(\lambda) &= \frac{\sin 2\eta}{\cosh(\lambda - i \eta)\cosh(\lambda+ i \eta)}, \\
	K_2(\lambda) &= \frac{\sin 4\eta}{\sinh(\lambda - 2 i \eta)\sinh(\lambda + 2 i \eta)}.
\end{align}

The representation of the wave-function in the Bethe Ansatz solvable systems is specifically convenient for computation of the correlation functions $\enn{m}$. It follows directly that
\begin{equation}
  \enn{N} = \modsq{\chi({\rm O}_N|\bflambda_M)}\cdot\modsq{\chi({\rm E}_N|\bflambda_M)},
\end{equation}
whereas a lower order correlation function is 
\begin{align}
  &\enn{N-2} = \Big|\chi^*({\rm O}_{N-2}, N| \bflambda_M)\chi({\rm E}_{N-2}, N|\bflambda_M) \nonumber\\
  &+ \chi^*({\rm O}_{N-2}, N\!-\!1| \bflambda_M)\chi({\rm E}_{N-2}, N\!-\!1|\bflambda_M)\Big|^2.
\end{align}
Here ${\rm E}_m = \{2, 4, \dots, m\}$ and ${\rm O}_m = \{1, 3, \dots, m-1\}$.

\section{Correlations in the Majumdar-Ghosh model} \label{app:MG}

In this Appendix we derive results for the correlations in the Majumdar-Ghosh model which we use in the main text. W consider a chain of $N$ sites with $N$ even and periodic boundary conditions. 
The ground state $\ket{\psi} =\mathcal N^{-1}(\ket{\psi_1} + \ket{\psi_2})$ of the Majumdar-Ghosh model is a superposition of two (normalized) states~\cite{Majumdar_Ghosh,Majumdar_1970} 
\begin{align}
	    \ket{\psi_1}&= \bigotimes_{j=0}^{\frac N2-1}\frac1{\sqrt2}\left(\ket{\uparrow_{2j+1},\downarrow_{2j+2}}-\ket{\downarrow_{2j+1},\uparrow_{2j+2}}\right)\\
    \ket{\psi_2}&=\bigotimes_{j=0}^{\frac N2-1}\frac1{\sqrt2}\left(\ket{\uparrow_{2j+2},\downarrow_{2j+3}}-\ket{\downarrow_{2j+2},\uparrow_{2j+3}}\right).
\end{align}
Norm of the ground state is then 
\begin{equation}
  \mathcal N^2=2^{-N/2+2}\left( 1 + 2^{N/2 - 1}\right),
\end{equation}
for the two states $\ket{\psi_i}$ are not orthogonal.
We use the following notation for the antiferromagnetic correlator for even $m$,
\begin{equation}
  \hat{\mathcal A}_m = \hat\sigma_+^{(1)}\hat\sigma_-^{(2)} \cdots \hat\sigma_+^{(m-1)}\hat\sigma_-^{(m)},
\end{equation}
This gives
\begin{subequations}
  \begin{align}
    \bra{\psi_{1,2}}\hat{\mathcal A}_N\ket{\psi_{1,2}} &= \left( -\frac{1}{2}\right)^{N/2}, \\
    \bra{\psi_{2,1}}\hat{\mathcal A}_N\ket{\psi_{1,2}} &=  \left( \frac{1}{2}\right)^{N/2}.
  \end{align}
\end{subequations}
which leads to the correlator
\begin{equation}
  \bra{\psi}\hat{\mathcal A}_N\ket{\psi}=\frac{1 + (-1)^{N/2}}{2} \frac{1}{1 + 2^{N/2-1}},
\end{equation}
yielding, for $N/2$ even, the $\enn{N}$ reported in the main text. We also compute $\enn{N-2}$ (as $\enn{N-1} = 0$). To this end we observe that
\begin{subequations}
  \begin{align}
    \bra{\psi_{1}}\hat{\mathcal A}_{N-2}\ket{\psi_{1}} &= \left(-\frac{1}{2}\right)^{N/2-1}, \\
    \bra{\psi_{2}}\hat{\mathcal A}_{N-2}\ket{\psi_{1}} &= - \left( \frac{1}{2} \right)^{N/2}, \\
    \bra{\psi_{2}}\hat{\mathcal A}_{N-2}\ket{\psi_{2}} &= 0.
  \end{align}
\end{subequations}
Therefore
\begin{equation}
  \bra{\psi}\hat{\mathcal A}_{N-2}\ket{\psi}= - \frac{1}{2} \frac{1}{1 + 2^{N/2 - 1}}.
\end{equation}

\end{document}